\documentclass[aps,twocolumn,prl,superscriptaddress,showpacs,tightenlines]{revtex4-1}
\usepackage{amsmath}
\usepackage{graphicx}
\usepackage{epsfig}
\usepackage{amsfonts}
\usepackage{color}

\begin{document}

\newcommand{\nn}{\nonumber} 
\newcommand{\ms}[1]{\mbox{\scriptsize #1}}
\newcommand{\msi}[1]{\mbox{\scriptsize\textit{#1}}}
\newcommand{\dg}{^\dagger}
\newcommand{\smallfrac}[2]{\mbox{$\frac{#1}{#2}$}}
\newcommand{\ket}[1]{| {#1} \ra}
\newcommand{\bra}[1]{\la {#1} |}
\newcommand{\pfpx}[2]{\frac{\partial #1}{\partial #2}}
\newcommand{\dfdx}[2]{\frac{d #1}{d #2}}
\newcommand{\half}{\smallfrac{1}{2}}
\newcommand{\s}{{\mathcal S}}
\newcommand{\jord}{\color{red}}
\newcommand{\kurt}{\color{blue}}

\title{The Energy Cost of Controlling Mesoscopic Quantum Systems}

\author{Jordan M.~Horowitz}
\affiliation{Department of Physics, University of Massachusetts at Boston, Boston, MA 02125, USA}
\author{Kurt Jacobs}
\affiliation{Department of Physics, University of Massachusetts at Boston, Boston, MA 02125, USA}
\affiliation{U.S. Army Research Laboratory, Computational and Information Sciences Directorate, ATTN: CIH-N, Adelphi, Maryland 20783, USA}
\affiliation{Hearne Institute for Theoretical Physics, Louisiana State University, Baton Rouge, LA 70803, USA}

\begin{abstract} 
We determine the minimum energy required to control the evolution of any mesoscopic quantum system in the presence of arbitrary Markovian noise processes. This result provides the mesoscopic equivalent of the fundamental cost of refrigeration, sets the minimum power consumption of mesoscopic devices that operate out of equilibrium, and allows one to calculate the efficiency of any control protocol, whether it be open-loop or feedback control. 
As examples we calculate the energy cost of maintaining a qubit in the ground state, the efficiency of resolved-sideband cooling of nano-mechanical resonators, and discuss the energy cost of quantum information processing.  
\end{abstract} 

\pacs{03.67.-a, 03.65.Yz, 05.70.Ln, 05.40.Ca}
\maketitle

Recent advances in the fabrication and control of mescoscopic quantum devices~\cite{Palomaki13, Mamin13, Safavi14, Silverstone14, Leghtas15} has made their potential application in future technologies evermore promising~\cite{Kelly15, Matthews13, Komar13}.
In such applications mesoscopic systems must be controlled to reduce the effects of environmental noise~\cite{WM10, Jacobs14}. Since reducing noise necessarily involves reducing the entropy of the controlled system, Landauer's principle suggests that there is an energetic cost, meaning that work must be supplied that can never be recovered. This energy cost is a fundamental question in quantum control and technologically important as it quantifies both the minimum power consumption and the minimum heat dissipation that must be handled by mescopic devices. Here we show that it is possible to fully characterize, in a relatively simple way, the minimum power required for continuous control of any mesoscopic quantum system subjected to arbitrary Markovian noise~\footnote{While we restrict ourselves here to Markovian noise processes, we suspect that our results can be extended to non-Markovian open systems, and this is an interesting question for further investigation.}. 


There is a natural division of controlled systems into \textit{weakly coupled} and \textit{strongly coupled}, depending on how large their interaction with the controller. For weakly-coupled systems --  which include most present-day mesoscopic systems~\footnote{Weakly-coupled systems include superconducting nano-electromechanical circuits~\cite{Palomaki13, Kelly15}, NV-centers in diamond~\cite{Mamin13}, cavity-QED~\cite{ Komar13}, and photonic circuits~\cite{Silverstone14}.} -- the coupling does not appreciably change  the system's energy levels. As a result, the control does not affect the noise processes perturbing the system, but only adds Hamiltonian terms to the dynamics that facilitate control. For strongly-coupled systems the coupling does modify the system's energy levels and with that the environmental noise. It means that the controlled system and controller cannot be treated as thermodynamically separate. 


\emph{Preliminaries.---} The evolution of a mesoscopic system $\s$ weakly-coupled to its surroundings is given by a linear differential equation for its density matrix $\rho$.
Denoting the Hamiltonian of the system as $H_{\mathcal S}$ and the linear super-operators that model the irreversible dynamics induced by the environemental noise processes ${\mathcal D}_{\mathcal S}^i$, $i=1,\dots,N$, the equation of motion for $\s$ in the absence of any control mechanism is the Lindblad master equation 
$\dot\rho = -(i/\hbar)[H_{\mathcal S},\rho]+\sum_i{\mathcal D}_{\mathcal S}^i(\rho)\equiv{\mathcal L}_{\mathcal S}(\rho)$~\cite{Breuer07, Jacobs14}. We further assume each noise process has an invariant state $\pi^i$, given as ${\mathcal D}_{\mathcal S}^i(\pi^i)=0$.
For example, noise from a thermal reservoir at temperature $T$, would have as a fixed point the Boltzmann density $\pi^{\rm eq} \propto e^{-H_{\mathcal S}/T}$ (with $k_{\rm B}=1$, assumed throughout). Thus, we can view the overall dynamics as a competition between noise processes, each trying to impose its own steady state onto the system.
The net effect is that in the absence of control, ${\mathcal S}$ will relax to a noise-induced steady-state density matrix $\rho^{\rm ss}$, given as the solution of ${\mathcal L}_{\mathcal S}(\rho^{\rm ss})=0$. The goal of control is to maintain $\s$ in an arbitrary state $\rho^*\neq \rho^{\rm ss}$.
\begin{figure}[b]
\includegraphics[width=1\hsize]{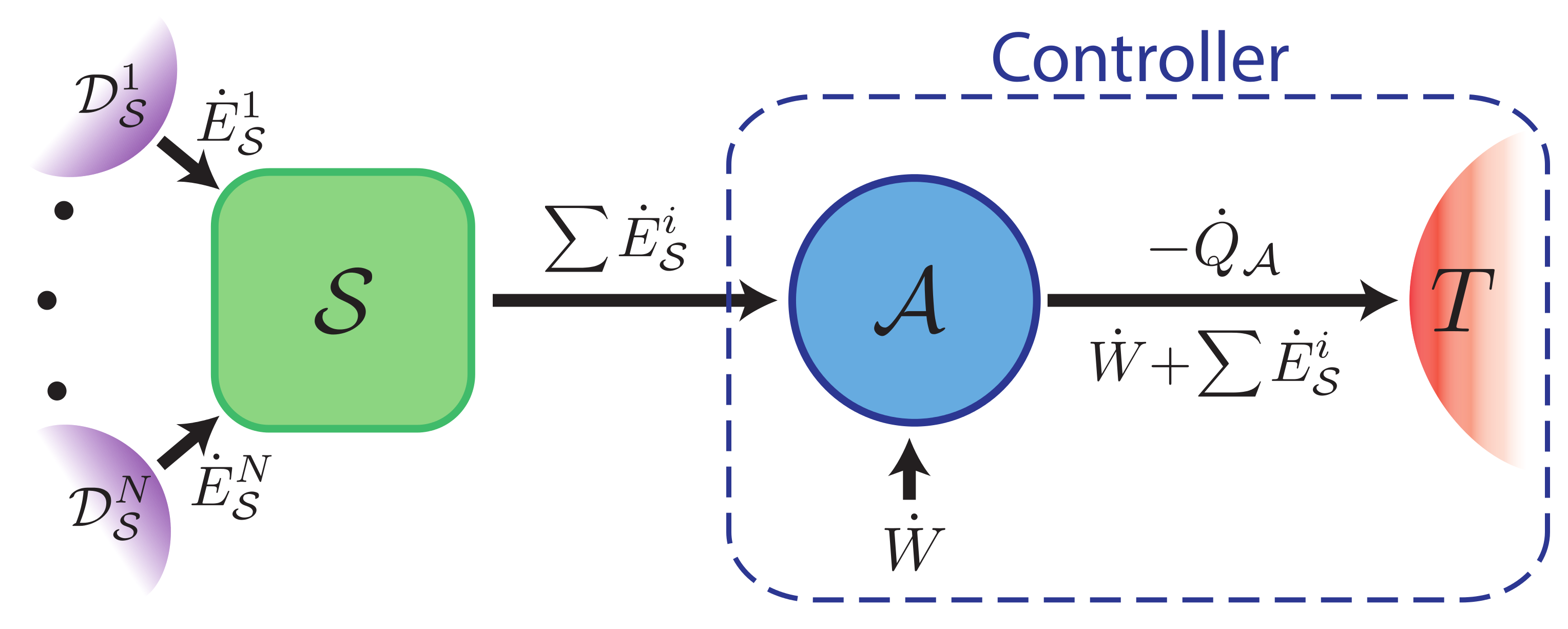}
\caption{Diagram of energy flow in a control process:  A system ${\mathcal S}$ (green square) bombard by noises $\{{\mathcal D}_{\mathcal S}^i\}$ is manipulated by a control system, consisting of an auxiliary ${\mathcal A}$ (blue circle) that extracts noise and energy $\sum E^i_{\mathcal S}$ from ${\mathcal S}$, depositing this energy as heat $-{\dot Q}_{\mathcal A}$ in its environment by supplying work ${\dot W}$.  Time's arrow is evident by the directional flow of energy through the figure from left to right, which demands that energy is dissipated.}
\label{fig:cartoon}
\end{figure}

{\it Weakly-coupled control.---} 
Control is implemented by weakly coupling $\s$ to an auxiliary quantum system ${\mathcal A}$ immersed in a thermal bath at temperature $T$, as in Fig.~(\ref{fig:cartoon}),  in such a way that the reduced steady-state of ${\mathcal S}$ is $\rho^*$.   
The assumption that $\s$ and ${\mathcal A}$ are weakly coupled guarantees that the dynamics induced  in $\s$ by its surroundings, given by $\{{\mathcal D}_{\mathcal S}^i\}$, is not changed by the coupling; yet, the control system can still affect $\s$'s evolution.
Thus, the evolution of the joint density matrix $\tau$ of $\s\oplus{\mathcal A}$ can be modeled as 
\begin{equation}\label{eq:master}
{\dot \tau} = (-i/\hbar)[{\mathcal H}(t),\tau]+\sum_i{\mathcal D}_\s^i(\tau)+{\mathcal D}_{\mathcal A}(\tau),
\end{equation}
in terms of the time-dependent joint Hamiltonian ${\mathcal H}(t)=H_\s+H_{\mathcal A}+V(t)$ with auxilliary Hamiltonian $H_{\mathcal A}$ and weak interaction $V(t)\ll H_\s, H_{\mathcal A}$, and the thermal-noise operator affecting the auxiliary, ${\mathcal D}_{\mathcal A}$. It is important to note that while we perform our analysis by coupling the system to a mesoscopic auxiliary system, the results apply to any control method, since this scenario has measurement-based control as a special case~\cite{Jacobs14b, Jacobs14}. 

Our main result is the minimum power ${\dot W}$ that the controller must supply to control $\s$, which crucially depends only on $\s$'s surroundings and the target state $\rho^*$. We prove this result rigorously -- the proof is outlined below, and is detailed in the Supplemental material~\footnote{See Supplemental Material at XXXXX for further details of the proofs of Eqs.~(\ref{eq:Wmin}) and (\ref{eq:strong}), the nonadiabatic entropy production, and the definition of weak-coupling.} -- but it can be understood in terms of an intuitive picture. It is due to the fact that for any isothermal process the work done on a system is bounded by the relation $W\ge \Delta F$, where $\Delta F$ is the change in the nonequilibrium free energy $F(\rho)=E(\rho)-TS(\rho)$, with average energy $E(\rho)={\rm Tr}[\rho H_{\s}]$ and von Neumann entropy $S=-{\rm Tr}[\rho\ln\rho]$~\cite{Esposito11}. With this in mind, each noise source ${\mathcal D}^i_\s$ continuously pushes the state of $\s$ away from $\rho^*$, and in doing so changes its free energy, implying that the controller must supply a commensurate amount of work to restore this free energy. Specifically, in the controlled steady state, the noise perturbations are changing $\s$'s entropy at a rate ${\dot S}^i_\s(\rho^*) = - {\rm Tr}[{\mathcal D}^i_{\mathcal S}(\rho^*)\ln\rho^*]$ while pumping energy in at a rate ${\dot E}_\s^i(\rho^*)= {\rm Tr}[{\mathcal D}^i_{\mathcal S}(\rho^*) H_\s]$. To undo these perturbations the controller must continuously transfer this entropy and energy through ${\mathcal A}$, eventually dumping it in ${\mathcal A}$'s thermal reservoir at temperature $T$. We show that this requires a minimum work rate
\begin{align}\label{eq:Wmin}  
{\dot W}_{\rm min} &= -\sum_i\dot{F}^i_{\mathcal S}(\rho^*) = \sum_i T{\dot S}^i_{\mathcal S}(\rho^*)-{\dot E}_{\mathcal S}^i(\rho^*)  \nonumber \\
&=-\sum_i {\rm Tr}[{\mathcal D}_\s^i(\rho^*)(T\ln \rho^* + H_\s)],
\end{align} 
where ${\dot F}^i_{\s}$ is the rate of change of the free energy of $\s$ due to the noise, evaluated at the temperature of the {\it auxiliary's} thermal reservoir.
To summarize, the noise affects the free energy of $\s$, and the controller must undo this change in free energy, requiring work; the reference temperature is that of ${\mathcal A}$'s thermal reservoir, since the energy is ultimately dissipated there.
This bound is for the energetics of the joint system, and therefore not a manifestation of the nonadiabatic entropy production~\cite{Yukawa:2001tf,Sagawa:2012vd,Gardas2015} for quantum nonequilibrium steady states (see~\cite{Note3}).
We now explore the consequences of Eq.~(\ref{eq:Wmin}). 

First, notice that ${\dot W}_{\rm min}$ may be negative, meaning that we can extract energy while controlling the system. For example, when the system is coupled to a hot bath at $T_H$ and a cold one at $T_C$, our target state $\rho^*$ may coincide with the regime where $\s$  operates as a heat engine. However, for isothermal control in which $\s$ sees a single bath at temperature $T$, ${\dot W}_{\rm min}$ must be positive as required by the second law. Another important scenario is that of maintaining $\s$ in a pure (zero-entropy) state. Since the derivative of entropy at zero is infinity, such control requires an infinite rate of work as reflected by the term $\ln \rho$ in Eq.~(\ref{eq:Wmin}). It is for the same reason that the power required for macroscopic refrigeration tends to infinity as the cold temperature tends to zero. Finally, our result also supplies the minimum work to push the system through a specified sequence of states $\rho^*(t)$, from $t=0$ to $\theta$, since the energy cost at any particular time depends only the system's state at that time: ${\dot W}_{\rm min} = -\int_0^\theta \sum {\dot F}^i[\rho^*(t)]\,dt$. Via this bound one can quantify the energetic efficiency of finite-time protocols such as shortcuts to adiabaticity \cite{Deffner2014}. 

\begin{figure}[t] 
\includegraphics[width=1\hsize]{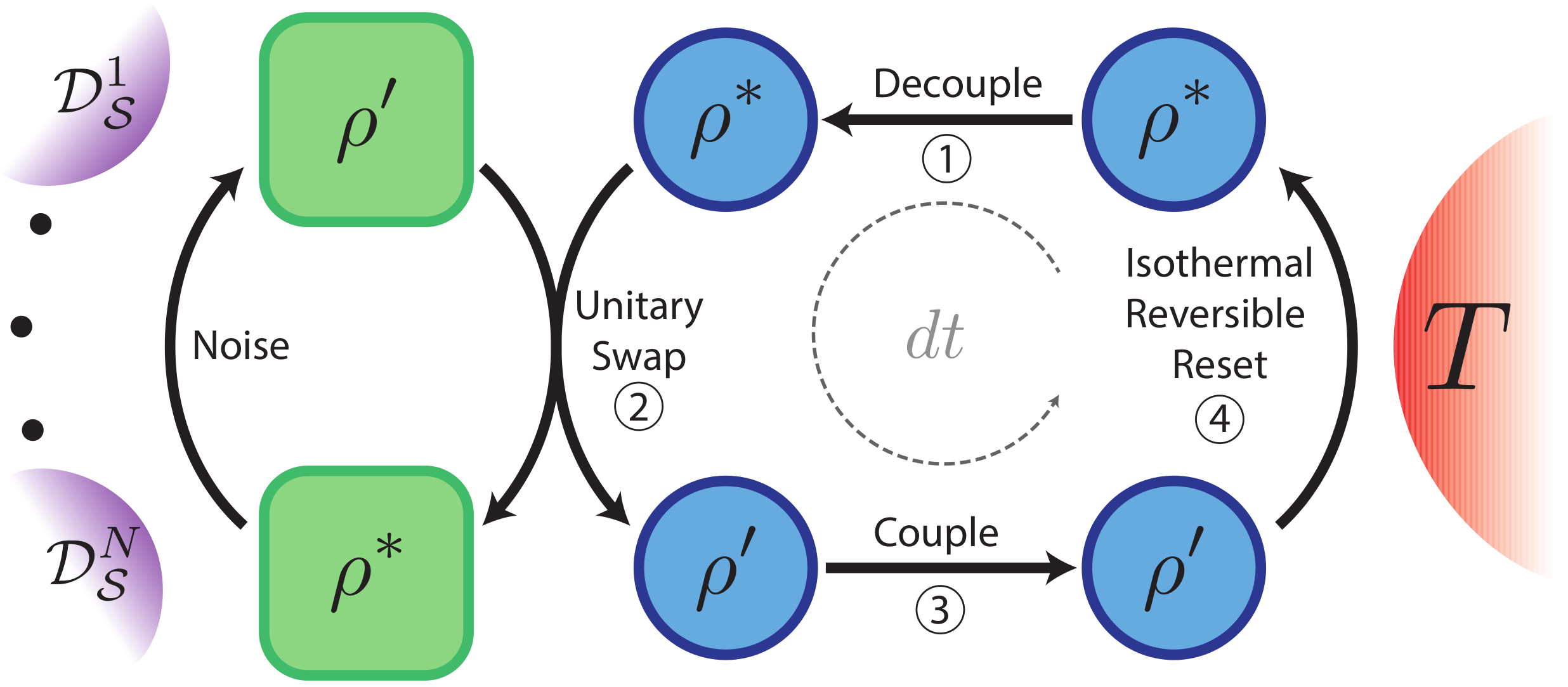}
\caption{Nonautonomous optimal control protocol: In a small interval of time $dt$, noise perturbs the state of $\s$ (green square) from $\rho^*\to\rho^\prime = \rho^*+{\mathcal L}_{\mathcal S}(\rho^*)dt$.  An optimal controller is formed by an auxiliary system ${\mathcal A}$ (blue circle) whose Hamiltonian can be set to match $\s$'s.  At the beginning of the time interval, ${\mathcal A}$'s state is the target state $\rho^*$, and control proceeds in 4 steps: 1) ${\mathcal A}$ is decoupled from its thermal environment, 2) A unitary swap is performed on $\s\oplus{\mathcal A}$ changing ${\mathcal A}$'s state to $\rho^\prime$, 3) ${\mathcal A}$ is coupled to its thermal environment, and 4) ${\mathcal A}$ is isothermally and reversibly reset to $\rho^*$, doing work $W = F(\rho^*)-F(\rho^\prime)\approx- \sum {\dot F}^i_{\s}(\rho^*)dt$.}
\label{fig:optimal}
\end{figure}

Our analysis further reveals that the minimum work, Eq.~(\ref{eq:Wmin}), can be achieved when the auxiliary operates reversibly. This requires a separation of time-scales, where the thermal relaxation of the auxiliary is very fast allowing it to remain essentially always in equilibrium. Additionally, the auxiliary's Hamiltonian dynamics must be fast compared to the system dynamics in order to rapidly extract the noise. An explicit nonautonomous protocol that implements this time-scale separation is described in Fig.~(\ref{fig:optimal}), where the rapid auxiliary dynamics are exploited to complete a reversible control cycle in every infinitesimal moment of time.

{\it Strongly-coupled control.---} When the auxiliary is coupled to $\s$ so strongly that it changes the energy levels of $\s$, it also changes the effect of the environment on $\s$ by altering the $\{{\mathcal D}_\s^i\}$. Because of this we can no longer bound the minimum work to control $\s$ solely in terms of the properties of $\s$, because the result will depend on the choice of joint Hamiltonian ${\mathcal H}$ through the interaction.  

First we observe that if we have access to any joint Hamiltonian ${\mathcal H}$, no work is required to sustain $\s$ in an arbitrary constant state $\rho^*$: We can always choose a fixed ${\mathcal H}$ so that the energy levels and eigenstates of $\s$ set $\rho^{\rm ss} = \rho^*$. Since ${\mathcal H}$ is time-independent, no work is required. Therefore the problem is well-motivated only when the interaction $V$ is restricted to a subset $\mathcal{V}$ of all interactions. We include in $\mathcal{V}$ all weak-coupling Hamiltonians, defined as those that do not change appreciably the eigenstates or eigenvalues of the system. This allows any control of the system that is slow compared to its dynamics, and unlimited control of the auxiliary~\cite{Note3}. 

We show that the minimum work to control strongly coupled systems is
\begin{equation}\label{eq:strong}
{\dot W}_{\rm min}=\min_{{\mathcal H},\tau} \left[-\sum_i {\rm Tr}[{\mathcal D}^i_{\mathcal S}(\tau)(T\ln \tau+ {\mathcal H})]\right],
\end{equation}
where the minimum is taken over all ${\mathcal H}\in{\mathcal V}$ and $\tau$ such that ${\rm Tr}_{\mathcal A}[\tau]=\rho^*$. The proof is an extension of that for Eq.~(\ref{eq:Wmin})~\cite{Note3}. We show this bound is tight by demonstrating a protocol that saturates it, akin to the weak coupling protocol in Fig.~(\ref{fig:optimal}). Suppose we know the Hamiltonian and density matrix that gives the minimum in Eq.~(\ref{eq:strong}): call them ${\mathcal H}_{\rm m}$ and $\tau_{\rm m}$. Then in a small time interval $dt$ the system's noise perturbs the joint system, causing an evolution $\tau_{\rm m}\to\tau^\prime$. To undue these perturbations, we couple the joint system to ${\mathcal A}$'s thermal reservoir and rapidly and reversibly raise ${\mathcal A}$'s energy levels, so that the joint state becomes $\sigma\otimes|0\rangle\langle0|$, for some system state $\sigma$. The auxiliary is then uncoupled from the bath, and the state of the system is swapped into the auxiliary: $|0\rangle\langle0|\otimes\sigma$. This will usually require a weak interaction, which is allowed since it is only strong changes to $V$ that are constrained. Now that all nontrivial structure of the state is contained in the auxiliary, we can use our ability to arbitrarily manipulate the auxiliary Hamiltonian to isothermally and reversibly return the joint state to the initial state $\tau_{\rm m}$. Since the entire process that takes the joint state from $\tau^\prime\to\tau_{\rm m}$ is reversible, the work equals the free energy difference $\Delta F = F(\tau_{\rm m}) - F(\tau^\prime)$ predicted in Eq.~(\ref{eq:strong}). Note that here we have restricted ourselves to a constant or slowly varying $\rho^*$, and thus ${\mathcal H}_{\rm m}$, to ensure that the system evolution is Markovian.
 
\textit{Resolved-sideband cooling.---} Resolved-sideband cooling is the current state-of-the-art in cooling mechanical quantum resonators~\cite{Wilson-Rae07, Marquardt07, Schliesser09, Tian09, Wang11} and is an example of coherent feedback control~\cite{Hamerly12, Jacobs15}. The auxiliary system is an optical or superconducting oscillator with a frequency, $\Omega$, sufficiently high that it sits in its ground state at the ambient temperature $T$. Cooling is accomplished by coupling the oscillators linearly and modulating this coupling at the frequency difference. In the weak coupling (rotating-wave) approximation the interaction Hamiltonian is $V = G + G^\dagger$ with $G = \hbar g a^\dag b e^{-i(\Omega-\omega)t}$; $\omega$ is the mechanical frequency, and $a$ and $b$ are the respective annihilation operators for the oscillators. This driven interaction mediates quanta exchange between the resonators, with the energy difference per quantum supplied as work $\Delta w=\hbar(\Omega-\omega)$. To achieve cooling the auxiliary must dump energy into the bath with sufficient speed.

\begin{figure}
\includegraphics[width=1\hsize]{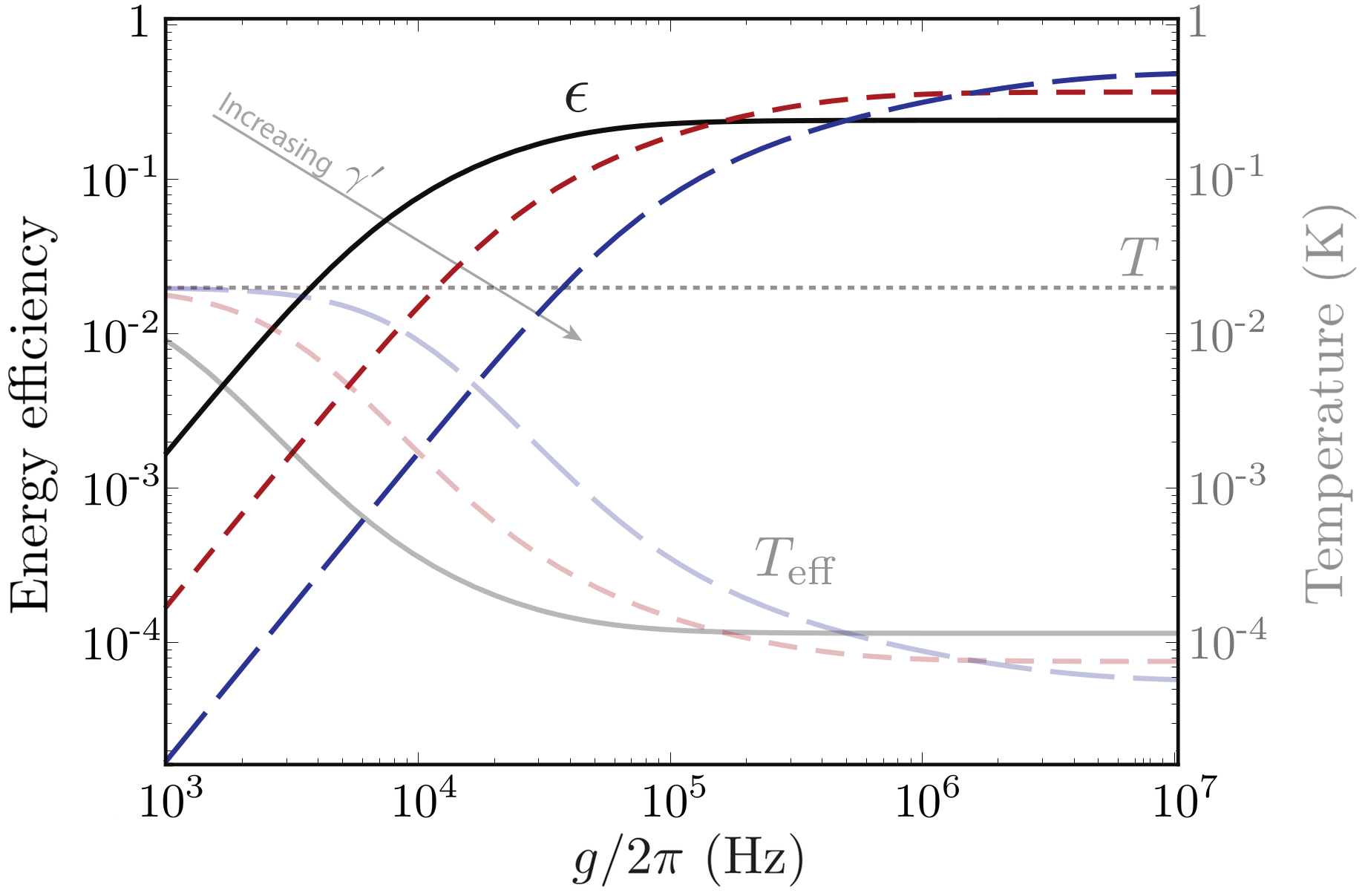}
\caption{Plot of the efficiency $\epsilon$ of resolved-sideband cooling of a mechanical oscillator in the weak-coupling regime, as a function of the interaction rate $g$ for three values of the auxiliary oscillator's damping rate, $\gamma^\prime/2\pi = 10^5~{\rm Hz}$ (solid black), $10^6~{\rm Hz}$ (dashed red), and $10^7~{\rm Hz}$ (long-dashed blue). Corresponding opaque curves give the cooled effective temperature $T_{\rm eff}$, and the grey dotted line is the ambient temperature $T$. Parameters from~\cite{Teufel11} are $\omega/2\pi = 10.56~{\rm MHz}$, $\Omega/2\pi = 1.54~{\rm GHz}$, $T = 20~{\rm mK}$, and $\gamma/2\pi = 32~{\rm Hz}$.} 
\label{fig:eff}
\end{figure}

Due to the linear dynamics the cooled steady state of the mechanical oscillator under sideband cooling is a Boltzmann-like equilibrium state at an effective temperature, $T_{\rm eff}<T$. While $T_{\rm eff}$ is not a true temperature we will see that it is useful. The rate at which the bath increases the oscillator entropy in the cooled state is ${\dot S}_{\s} =\dot Q_{\s} / T_{\rm eff}$, with ${\dot Q}_{\s}$ the heat flow from bath into the oscillator and equivalently the energy flow to the auxiliary. Thus rather strikingly the mesoscopic oscillator has an entropy production rate identical to that of a  thermal bath at temperature $T_{\rm eff}$. 
Because of this the cooling efficiency has precisely the form of that of a macroscopic refrigerator. Defining the coefficient of performance in the usual way as $\eta^{\rm COP}={\dot Q}_{\s}/{\dot W}$, where ${\dot W}$ is the actual power consumed by the fridge~\cite{Callen}, the minimum power can be written as ${\dot W}_{\rm min} = \dot{Q}_{\s} \eta^{\rm COP}_{\rm ideal}$, with efficiency 
\begin{equation}\label{eq:sideEff}
\epsilon = {\dot W}_{\rm min} /{\dot W} = \eta^{\rm COP} /\eta^{\rm COP}_{\rm ideal}. 
\end{equation}
Here $\eta^{\rm COP}_{\rm ideal}=T/T_{\rm eff}-1$ is the ideal Carnot coefficient of performance. In Fig.~(\ref{fig:eff}), we plot the effective temperature and efficiency achieved by sideband cooling as a function of the interaction rate $g$, using parameters from the recent experiment in~\cite{Teufel11}. We see that stronger coupling gives increased efficiency and a colder temperature, and that high damping is only effective when the coupling transfers the entropy with sufficient speed. 

\textit{Cooling a single qubit.---} The master equation describing a weakly-damped qubit with energy gap $E$ in contact with a bath at temperature $T$ can be found in~\cite{Breuer07, Jacobs14}. We wish to maintain the qubit at a temperature $T_{\ms{c}} < T$. If $T_{\ms{c}}/E \ll 1$, so that the qubit is close to its ground state, the required power from Eq.~(\ref{eq:Wmin}) is simply 
\begin{equation}
    {\dot W}_{\rm min} \approx \gamma n_T  E \left( T/T_{\ms{c}} - 1 \right) , \label{qbcool} 
\end{equation}
with $\gamma$ the qubit's damping rate and $n_T = 1/(e^{E/T} - 1)$. The power goes to infinity as $T_{\ms{c}}\rightarrow 0$ as expected. The full expression for arbitrary $T_{\ms{c}}$ is obtained by replacing $n_T$ by the $(z-w)/[(z-1)(w+1)]$ where $z=\exp(E/T)$ and $w=\exp(E/T_{\ms{c}})$.  
  
Coupling the qubit strongly to an auxiliary qubit with an energy gap $\mathcal{E} > E$ provides a simple example in which a strong interaction reduces the power requirements for ground-state cooling. The interaction allows us to effectively increase the energy gap of the first qubit, increasing the equilibrium population of the ground state and thus reducing the effort required to preserve that state. Let the auxiliary gap be $\mathcal{E} \gg kT$ so that it effectively sits in its ground state $|0\rangle$, and take the Hamiltonian of the two qubits as $H = E \sigma_z^{(1)}/2 + H_{\ms{I}} + \mathcal{E} \sigma_z^{(2)}/2$ with the interaction $H_{\ms{I}} = g \sigma_z^{(1)} \sigma_z^{(2)}$/2. Since the auxiliary is in state $|0\rangle$, if we set $g = -\varepsilon$, and assuming that $\mathcal{E} > \varepsilon > E$, then the two lowest energy states of the joint system are $|0\rangle |0\rangle$ and $|1\rangle |0\rangle$, where $|i\rangle|j\rangle$ denotes system state $|i\rangle$ and auxiliary state $|j\rangle$. These two states have an energy gap of $E + \varepsilon$, so the interaction effectively increases the energy gap of the first qubit by $\varepsilon$. The minimum power consumption is then given by replacing $E$ with $E+\varepsilon$ in Eq.~(\ref{qbcool}). 

\textit{Devices that operate out-of-equilibrium.---} A quantum computer is one such device. While quantum logic gates are unitary and thus require no energy, a quantum computer consumes power because the constituent qubits are subject to relaxation (errors) from environmental noise. The error-correction process continually introduces new qubits prepared in near-pure states to combat these errors~\cite{Gottesman09, Aliferis06, Knill05}. We can estimate the energy consumption per qubit for a quantum computer by using a simple error model, and averaging the minimum energy dissipation for a single qubit over all pure states. Since fault-tolerant computation requires that the qubits are refreshed while the errors are still small, the analysis we have performed above for continuous-time control is appropriate. However we restrict to logic gates that are slow compared to the qubit frequency to ensure the damping is Markovian~\cite{Alicki06}. A typical error model involves thermal damping at (effectively) zero temperature at rate $\gamma$, and depolarizing at rate $\beta$ for which the master equation is $\dot{\rho} = -(\gamma/2) (\sigma^\dagger \sigma \rho + \rho \sigma^\dagger \sigma - 2 \sigma \rho \sigma^\dagger ) - (\beta/4)\sum_j [\sigma_j, [\sigma_j, \rho]]$ with $j=x,y,z$. The change in free energy averaged over all pure states is $\Delta F \approx (p_{\beta} \ln p_{\beta} +  p_{\gamma} \ln p_{\gamma})kT  - p_{\gamma} E$, where $p_{\beta} = \beta \tau \ll 1$ and $p_{\gamma} = \gamma \tau/2 \ll 1$ are the error probabilities due to the thermal damping and depolarizing, respectively, and $E$ is the energy gap of the qubit. The time $\tau$ is the duration of a single lowest-level fault-tolerant gate, which includes the auxiliary qubits injected for error-correction and/or teleportation operations, both or which refresh the working qubits~\cite{Gottesman09}. The minimum energy consumption of a computation is therefore $M \Delta F$ where $M$ is the total number of qubits injected during the computation. Given the above form of $\Delta F$, we can conclude that if quantum computers run with $kT \ll E$ as presently envisaged, the minimum energy cost will be dominated by the loss of the qubits' internal energy to the bath. 

{\it Outline of proofs of Eqs.~(\ref{eq:Wmin}) and (\ref{eq:strong}).---} The key ingredient is a second-law-like inequality for the entropy production of an open quantum system modeled with a Lindblad master equation, which follows from the monotonicity of the quantum relative entropy under Markovian noise~\cite{Spohn1978, Alicki04, Sagawa08}.  If we define $\Sigma_\s^{i} = -  {\rm Tr}[{\mathcal D}^i_\s(\tau)(\ln \tau-\ln \pi^i)]$ and $\Sigma_{\mathcal{A}} = - {\rm Tr}[{\mathcal D}_{\mathcal A}(\tau)(\ln \tau-\ln \pi^{\rm eq}_{\mathcal A})]$, then $\Sigma = \sum_i \Sigma_\s^i + \Sigma_{\mathcal{A}}$ gives the total entropy production of the joint system. Further, $\Sigma_{\mathcal{A}}$ and the $\Sigma_\s^i$ are time derivatives of relatives entropies under Markovian noise processes. As such they are negative and thus $\Sigma \geq 0$. If we drop all the entropy production due to ${\mathcal A}$ we obtain $\Sigma \ge \sum_i \Sigma_\s^i \ge 0$. We now trace over ${\mathcal A}$ because we want a bound purely in terms of $\s$. This operation decreases $\Sigma$ due to the monotonicity of the relative entropy under partial trace~\cite{Sagawa08}, giving us $\Sigma \ge - \sum {\rm Tr}[{\mathcal D}^i_\s(\rho^*)(\ln \rho^*-\ln \pi^i)] \ge 0$. We next note that in the steady state $\Sigma = -\sum {\rm Tr}[{\mathcal D}^i_{\mathcal S}(\rho^*)\ln\pi^i]-{\dot Q}_{\mathcal A}/T$, in terms of the heat flow ${\dot Q}_{\mathcal A}=-{\rm Tr}[{\mathcal D}_{\mathcal A}(\tau)\ln\pi^{\rm eq}_{\mathcal A}] $ out of ${\mathcal A}$'s reservoir. The relation in Eq.~(\ref{eq:Wmin}) then follows from energy conservation in the steady state, $-{\dot Q}_{\mathcal A}={\dot W}+\sum {\dot E}^i_\s$. To obtain Eq.~(\ref{eq:strong}) the steps are the same as those above, except that we minimize the right hand side of $\Sigma \ge \sum_i \Sigma_\s^i$ over the interaction $V$, and then skip the step in which we trace over ${\mathcal A}$~\cite{Note3}. 

\textit{Acknowledgments:} JH and KJ were partially supported by the ARO MURI grant W911NF-11-1-0268, and KJ by the NSF project PHY-1212413. 


%

\end{document}